\documentclass[a4paper,11pt]{article}

\usepackage{jheppub}

\usepackage[T1]{fontenc} 

\usepackage{graphicx}
\usepackage{epsfig}
\usepackage{rotating}
\usepackage{amssymb}
\usepackage{subfigure}
\usepackage{dsfont}
\usepackage{psfrag}
\usepackage{amsmath,euscript,array,mathrsfs}
\usepackage{axodraw}
\usepackage{bbold}
\usepackage{epsf}

\newcommand{\beq}{\begin{equation}}
\newcommand{\eeq}{\end{equation}}
\newcommand{\beqs}{\begin{eqnarray}}
\newcommand{\eeqs}{\end{eqnarray}}
\newcommand{\lsim}{\mathrel{\raisebox{-
.6ex}{$\stackrel{\textstyle<}{\sim}$}}}
\newcommand{\gsim}{\mathrel{\raisebox{-
.6ex}{$\stackrel{\textstyle>}{\sim}$}}}
\newcommand{\Tr}{{\rm Tr}}

\def\hbar{\hspace{0pt}\raisebox{1pt}{$-$} \hspace{-7pt} h}

\def\di{\mbox{d}}

\newcommand{\be}{\begin{equation}}
\newcommand{\ee}{\end{equation}}

\newcommand{\bea}{\begin{eqnarray}}
\newcommand{\eea}{\end{eqnarray}}

\def\lbldef#1#2{\expandafter\gdef\csname #1\endcsname {#2}}

\def\href#1#2{#2}

\newcommand{\ber}{\begin{eqnarray}}
\newcommand{\eer}{\end{eqnarray}}

\newcommand{\beqar}{\begin{eqnarray}}

\newcommand{\eeqar}{\end{eqnarray}}

\newcommand{\dsl}
  {\kern.06em\hbox{\raise.15ex\hbox{$/$}\kern-.56em\hbox{$\partial$}}}

\newcommand{\eeqarr}{\end{eqnarray}}
\newcommand{\ZZ}{{\rm \kern 0.275em Z \kern -0.92em Z}\;}

\def\CC{{\mathchoice
{\rm C\mkern-8mu\vrule height1.45ex depth-.05ex
width.05em\mkern9mu\kern-.05em}
{\rm C\mkern-8mu\vrule height1.45ex depth-.05ex
width.05em\mkern9mu\kern-.05em}
{\rm C\mkern-8mu\vrule height1ex depth-.07ex
width.035em\mkern9mu\kern-.035em}
{\rm C\mkern-8mu\vrule height.65ex depth-.1ex
width.025em\mkern8mu\kern-.025em}}}

\def\RR{{\rm I\kern-1.6pt {\rm R}}}

\def\ZZ{{\rm Z}\kern-3.8pt {\rm Z} \kern2pt}
\def\IB{\relax{\rm I\kern-.18em B}}
\def\ID{\relax{\rm I\kern-.18em D}}
\def\II{\relax{\rm I\kern-.18em I}}
\def\IP{\relax{\rm I\kern-.18em P}}

\newcommand{\bear}{\begin{eqnarray}}
\newcommand{\eear}{\end{eqnarray}}

\def\6{\partial}

\def\bea{\begin{eqnarray}}
\def\eea{\end{eqnarray}}

\def\beqx{\begin{displaymath}}
\def\eeqx{\end{displaymath}}

\newcommand{\bmat}{\left(\begin{array}}
\newcommand{\emat}{\end{array}\right)}

\def\bo{{\raise-.3ex\hbox{\large$\Box$}}}               % D'Alembertian
\def\face{{\raise.2ex\hbox{$\displaystyle \bigodot$}\mskip-2.2mu \llap {$\ddot
        \smile$}}}                                   % happy face
\def\>{\rangle}                                      %right angle
\def\<{\langle}                                      %left angle

\def\leftrightarrowfill{$\mathsurround=0pt \mathord\leftarrow \mkern-6mu
        \cleaders\hbox{$\mkern-2mu \mathord- \mkern-2mu$}\hfill
        \mkern-6mu \mathord\rightarrow$}        % <--> double differential
\def\dvec#1{\vbox{\ialign{##\crcr
        \leftrightarrowfill\crcr\noalign{\kern-1pt\nointerlineskip}
        $\hfil\displaystyle{#1}\hfil$\crcr}}}           % <--> accent
\def\Tr{{\rm Tr \,}}                                    % Trace
\def\-{\hphantom{-}}

\title{A spartan model for the LHC diphoton excess}

\author[a]{Thomas Appelquist,}

\affiliation[a]{Department of Physics, Sloane Laboratory, Yale University, New Haven, Connecticut 06520, USA}

\author[a]{James Ingoldby,}

\author[b]{ Maurizio Piai,}

\affiliation[b]{Department of Physics, College of Science, Swansea University,
Singleton Park, SA2 8PP, Swansea, Wales, UK}

 \author[a]{Jedidiah Thompson.}

\date{\today}

\abstract{
We propose a simple model accommodating the
reported $750$ GeV diphoton
excess seen in the first $13$-TeV run of the LHC. It leads to testable
predictions, in particular for di-lepton
production, at higher integrated luminosity. We append to the minimal
standard model a new gauge sector with
its own $SU(2)$ symmetry group. A new complex doublet scalar field
provides mass for the new vectors and
describes the $750$-GeV  resonance. An adequate rate for the diphoton
signals, with resonant production
via photon fusion, requires the VEV of the new scalar field to be
somewhat less than the electroweak scale.
This in turn requires the new heavy vectors to have sub-TeV masses and be relatively strongly
coupled. A new global $U(1)$ symmetry plays
a key role. Current precision-electroweak constraints are respected.
}

\begin{document}
\maketitle
\flushbottom

%%%%%%%%%%%%%%%%%%%%%%%%%%%%%%%%%%%%%%%%%%%%%%%%%%%%%%%%%
%%%%%%%%%%%%%%%%%%%%%%%%%%%%%%%%%%%%%%%%%%%%%%%%%%%

%%%%%%%%%%%%%%%%%%%%%%
\section{Introduction}
\label{Sec:intro}

Hints of new physics beyond the standard model (SM) continue to emerge from the LHC. From the 
recent run at $13$ TeV, the ATLAS and CMS collaborations have reported an excess in the diphoton mass
distribution at about $750$ GeV, with a statistical significance of $2-4 \sigma$~\cite{diphoton}.
The current data is compatible both with a very narrow resonance, and with a resonance 
of width ${\cal O}(50\,{\rm GeV})$.  Our analysis here favors the first option.
 We present a simple model, developed as an effective field theory 
 (EFT), which accommodates these signals,
is consistent with existing SM tests, and makes predictions that could be testable soon at the LHC.
The main ingredient  is an extension  of the SM gauge sector with the addition of a new, 
spontaneously broken $SU(2)$ gauge group.

In models with generic $SU(2)$ extensions of the SM gauge sector,
indirect bounds from precision measurements of electroweak and Higgs physics 
tend to favor the regime of  weakly-coupled
vectors in the multi-TeV mass range  (see for instance~\cite{ABIP} and references therein). 
Here we employ a simple model,  
designed to focus on the $750$-GeV diphoton signal and
associated physics, which allows for sub-TeV masses for the new vectors and is consistent with precision electroweak and Higgs measurements
(see also~\cite{CDD}).

We describe the Higgs and gauge sectors of the minimal SM using a complex doublet $\Phi$
with vacuum expectation value (VEV) $f$. 
We include a new $SU(2)_V$ gauge sector 
and a new complex-doublet field $\Phi_V$ with
VEV $f_V$  to provide mass for the three $SU(2)_V$
gauge bosons. 
The field $\Phi_V$ transforms as a bi-fundamental under $SU(2)_R \times SU(2)_V$, where the diagonal
$T^3$ generator of $SU(2)_R$ is electroweak gauged. 
After symmetry breaking, the residual massive scalar $H$ coming from $\Phi_V$
is our candidate for the  $750$ GeV resonance. The fermionic field content of the model
is that of the SM. The full field content of the model is shown in Table~\ref{Fig:3sites}.

With respect to other models in the recent literature~\cite{list}, ours
has a distinctive combination of  features. It has an accidental, approximate global $U(1)$ symmetry,
 which emerges together with a simple mechanism responsible for suppressing new contributions 
 to electroweak precision observables. Indirect bounds from electroweak and Higgs physics 
 are satisfied despite the presence of new sub-TeV vector and scalar particles.  
 The dominant production process for 
the new narrow heavy scalar particle $H$ is photon fusion~\cite{pf}. 
Since there are no new, fermionic degrees of
freedom,  the coupling of the heavy scalar particle $H$ to 
 photons is dominated by loops of the new vector bosons.

In Section~\ref{Sec:EFT} we describe the model in detail, and  in Section~\ref{Sec:precision} we discuss Higgs physics and
electroweak precision constraints. We discuss the properties of the $H$ resonance in
Section~\ref{Sec:H}, estimating its decay width and production cross section at the LHC. 
We conclude that
$f_V \lsim f$. 
In Section~\ref{Sec:V} we estimate the production and decays of the new, 
heavy vector states and conclude that the $SU(2)_V$ gauge coupling must be relatively strong to
make the vector states heavy enough to have avoided detection. 
Nevertheless, production estimates indicate that these states should be accessible in future LHC runs. We will comment on the predictive power of the EFT in the relevant range of parameters.
 In Section~\ref{Sec:symmetry} we describe the
symmetry properties of our model and the issue of fine tuning.
We summarize our results in Section~\ref{Sec:discussion} and describe possible extensions of our study.

%%%%%%%%%%%%%%%%%%%%%%%%%%%%%
\section{The Model}
\label{Sec:EFT}

\begin{table}
\begin{center}
\begin{tabular}{|c|c|c|c|c|c|}
\hline\hline
\multicolumn{6}{|c|}{{
\begin{picture}(0,55)(130,0)
\GCirc(70,26){10}{0}
\Line(80,26)(120,26)
\GCirc(130,26){10}{0.6}
\Line(140,26)(180,26)
\GCirc(190,26){10}{0}
\Text(100,38)[c]{ \Large $\Phi$}
\Text(160,38)[c]{ \Large $\Phi_V$}
\Text(70,6)[c]{ $SU(2)_L$}
\Text(130,6)[c]{ $U(1)_Y$}
\Text(190,6)[c]{ $SU(2)_{V}$}
\Text(70,46)[c]{ $SU(2)_L$}
\Text(130,46)[c]{ $SU(2)_{R}$}
\Text(190,46)[c]{ $SU(2)_{V}$}
\end{picture}
}}
\cr
\hline\hline
{\rm ~~~Fields~~~} &$SU(2)_L$ & $SU(2)_{R}$  & $SU(2)_{V}$  & $U(1)_{B-L}$ & $U(1)_{Y}$\cr
\hline\hline
$\Phi$ & $2$ & $\bar{2}$ & $1$ &  $0$ & $\pm{1}/{2}$ \cr
$\Phi_V$ & $1$ & $2$ & $\bar{2}$ &  $0$ &  $\mp{1}/{2}$\cr
\hline
$q_{L}$ & $2$ & $1$ & $1$ &  ${1}/{3}$ & ${1}/{6}$\cr
$q_{R}$ & $1$ & $2$ & $1$ &  ${1}/{3}$ & ${1}/{6}\pm{1}/{2}$\cr
$\ell_{L}$ & $2$ & $1$ & $1$ &  $-1$ & $-{1}/{2}$\cr
$\ell_{R}$ & $1$ & $2$ & $1$ &  $-1$ & $-{1}/{2}\pm{1}/{2}$\cr
\hline\hline
$W_{\mu}$ & $3$ & $1$ & $1$  & $0$ & $0$\cr
$V_{\mu}$ & $1$ & $1$ & $3$  & $0$ & $0$\cr
$B_{\mu}$ & $1$ & ${\ast}$ & $1$  & $0$ & $0$\cr
\hline\hline
\end{tabular}
\end{center}
\caption{The field content of the model. The figure shows a diagrammatic representation of the model,
with the global (top) and local (bottom) symmetries acting on the scalars $\Phi$ and $\Phi_V$ made manifest.
All fermions $q_i$ and $\ell_i$ are written as Weyl spinors, and all  vectors are written as $2\times 2$ matrices (see main text for details).}
\label{Fig:3sites}
\end{table}

The field content of our model is listed in Table~\ref{Fig:3sites}.
The hypercharge $Y$ of the matter fields is given by
$Y= T^3_R+\frac{1}{2}(B-L)$, where $T^3_R$ is the diagonal generator of $SU(2)_R$.
We also write $B_{\mu}\equiv B^3_{\mu}\left[T^3_R+\frac{1}{2}(B-L)\right]$,
while $W_{\mu}=W_{\mu}^aT^a$ and $V_{\mu}=V_{\mu}^aT^a$, so that all gauge-boson fields are written as
$2\times 2$ matrices. We normalize the $T^a$  generators of $SU(2)$ by $\Tr T^aT^b=\frac{1}{2}\delta^{ab}$.
The Lagrangian, including operators up to dimension-four, is
\beqs
{\cal L}_{}&=&\label{Eq:Lag}
+2g\Tr W^{\mu}J_{L\,\mu}\,-\,2g^{\prime}\Tr B^{\mu} J_{Y\,\mu}\\
&&-\,\frac{1}{2}\Tr {W}_{\mu\nu}{W}^{\mu\nu}
\,-\,\frac{1}{2}\Tr {V}_{\mu\nu}{V}^{\mu\nu}\,
\,-\,\frac{1}{2}\Tr {B}_{\mu\nu}{B}^{\mu\nu}\nonumber\\
&&+\,
 \frac{1}{4}\Tr |D\Phi|^2+ \frac{1}{4}\Tr |D\Phi_V|^2\,-\,V(\Phi_i)\nonumber\\
 &&-\frac{1}{\sqrt{2}}\bar{q_L}\Phi y_q q_R\,-\,\frac{1}{\sqrt{2}} \bar{\ell_L}\Phi y_{\ell} \ell_R\,,\nonumber
\eeqs
with $y_q$ and $y_{\ell}$ being $SU(2)_R$-breaking Yukawa matrices that will give rise to the masses of
the SM fermions in the familiar way.
The field strength tensors are defined so that the gauge bosons are canonically normalized.
We denote with $g$, $g^{\prime}$ and $g_V$ the three gauge couplings.
 $J_{L\mu}$ and $J_{Y\mu}$ are electroweak matter currents bilinear
in the SM fermion fields.
The covariant derivatives for the scalars are
\beqs
D_{\mu}\Phi &\equiv& \partial_{\mu}\ \Phi\,-\,i\left(g{W}_{\mu}\ \Phi - g^{\prime}\Phi{B}_{\mu}\right)\,,\nonumber\\
D_{\mu}\Phi_V &\equiv& \partial_{\mu}\ \Phi_V\,-\,i\left(g^{\prime}{B}_{\mu}\ \Phi_V - g_V\Phi_V{V}_{\mu}\right)\,.\nonumber
\eeqs

We write the potential with the same conventions as in~\cite{ABIP}:
\beqs
V(\Phi_i)&=&
\,+\,\frac{\lambda}{16}  \left(\frac{}{}\Tr\left[\frac{}{}\Phi\Phi^{\dagger}-f^2\,\mathbb{I}_2\right]\right)^2
\,+\,\frac{\lambda_V}{16}  \left(\frac{}{}\Tr\left[\frac{}{}\Phi_V\Phi_V^{\dagger}-f_V^2\,\mathbb{I}_2\right]\right)^2
\,.
\eeqs
We have omitted the dimension-four mixing term $\Tr(\Phi\Phi^{\dagger}) \Tr(\Phi_V\Phi_V^{\dagger}) = 2 \Tr(\Phi\Phi_V\Phi_V^{\dagger}\Phi^{\dagger})$.
The only communication between $\Phi$ and $\Phi_V$ then
 arises via the weakly-coupled gauge field $B_{\mu}$. 
  In Section~\ref{Sec:symmetry} we will justify the omission of the mixing term by observing that
 with a TeV-scale cutoff for our EFT, it will be
 generated at the loop level, but with a very small coefficient.
  Minimization of the potential yields the VEVs $\langle \Phi \rangle=f\,\mathbb{I}_2$ and $\langle \Phi_V \rangle=f_V\,\mathbb{I}_2$.

In the absence of the weakly-coupled gauge field $B_{\mu}$, the global $SU(2)_R$ transformation can act independently
on $\Phi$ and $\Phi_V$. In effect, it splits into two $SU(2)$'s. The gauging via $B_{\mu}$ couples these two, leaving
an extra $U(1)_V$ global symmetry in addition to the gauged symmetries. Among the gauged symmetries, the unbroken
$U(1)_Q$ of QED is generated by $Q=T^3_L+Y+T^3_V$.  The additional global  $U(1)_V$ can be taken to be
\beqs
U(1)_V:\,\,\, (\Phi_{V}(x), V_{\mu}(x)) \,\rightarrow\,e^{i\beta_V T^3} (\Phi_{V}(x),V_{\mu}(x)) e^{-i\beta_V T^3}\,,
\eeqs
where $\beta_V$ is a real number. The $U(1)_V$ symmetry, preserved even in the presence of a mixing term in the potential,
stabilizes the charged $V_{\mu}^{\pm}$. This symmetry can and will be broken,
allowing for decay of the charged vector. The breaking will arise from unknown physics at a
high scale $\Lambda_V$, and will be communicated to our fields by higher-dimension operators suppressed by powers of $\Lambda_V$.~\footnote{
Equivalently, one can think of symmetry breaking as arising in the UV completion via small symmetry breaking terms, without the need to introduce
a parametrically large scale $\Lambda_V$.
}
The neutral $V_{\mu}^0$ mixes with the $Z_{\mu}$ via the $B_{\mu}$ gauge interaction.

The $2\times 2$ mass matrix ${\cal M}^2_+$ for the charged vectors, in the $(W^{\mu},V^{\mu})$ basis, is
\beqs
{\cal M}^2_+&=&
\frac{1}{4}
\left(\begin{array}{cc}
g^2 f^2 & 0\cr
0 &  g_V^2 f_V^2\cr
\end{array}\right)\,.
\eeqs
Since there is no mixing in the charged sector, we have
\beqs
g_{\mathrm{SM}}&=&g\,,\,\,\,\,\,
v_W\,=\,f\,,
\eeqs
where $v_W\simeq 246$ GeV is the electroweak scale and $g_{\mathrm{SM}} \simeq 0.65$ is the $SU(2)_L$  coupling in the SM.

The $3\times 3$ mass matrix for the neutral vector bosons in the basis $(W^{\mu},  B^{\mu}, V^{\mu})$ is
\beqs
{\cal M}^2_0&=&
\frac{1}{4}
\left(\begin{array}{ccc}
g^2 f^2& -g g^{\prime} f^2 & 0\cr
-g g^{\prime} f^2 & g^{\prime\,2} (f^2+f_V^2)& - g^{\prime}g_{V} f_V^2\cr
0 & -g^{\prime}g_V f_V^2& g_V^2 f_V^2\cr
\end{array}\right)\,.
\eeqs
The diagonalization of this mass matrix yields a massless photon, along with the $Z_\mu$ and a heavy $V^0_\mu$ vector, with masses
\beqs
&&M^2_{Z,V^0}\,=\,\frac{1}{8} \bigg\{\frac{}{}f^2 \left(g^2+g^{\prime\,2}\right)\,+f_V^2
   \left(g^{\prime\,2}+g_V^2\right)\,\\
   &&\pm\,\sqrt{f^4
   \left(g^2+g^{\prime\,2}\right)^2-2 f^2 f_V^2 \left(g_V^2
   \left(g^2+g^{\prime\,2}\right)+g^{\prime\,2} (g-g^{\prime}) (g+g^{\prime})\right)+f_V^4
   \left(g^{\prime\,2}+g_V^2\right)^2}\bigg\}\,.\nonumber
\eeqs
To ensure that $M_{V^0}^2 \gg M_Z^2$, the term $f_V^{4}(g^{\prime\,2}+g_V^2)^2$ must dominate in the square root.
To describe the properties of the $750$ GeV scalar, 
we will find it necessary to keep $f_V \lsim f$, and therefore we will take $g_V^2 \gg  g^2, g^{\prime \, 2}$. In this limit,
\beqs
M_{Z}^2&\simeq & \frac{1}{4}(g^2+g^{\prime\,2})f^2\,-\,\frac{1}{4}\frac{g^{\prime\,4}}{g_V^2}f^2+\cdots\,,\\
M_{V^0}^2&\simeq&\frac{1}{4}(g_V^2+g^{\prime\,2})f_V^2\,+\,\frac{1}{4}\frac{g^{\prime\,4}}{g_V^2}f^2+\cdots\,.
\eeqs

The eigenstate of the photon $A_{\mu}$ is
\beqs
A_{\mu}&=&\frac{1}{\sqrt{g^2g_V^2+g^{\prime\,2}g^2+g^{\prime\,2}g_V^2}}\left(\frac{}{}g_Vg^{\prime}W_{\mu}^3\,+\,g_Vg B_{\mu} \,+\,g g^{\prime}V^3_{\mu}\right)\,.
\eeqs

The analogous expressions for the heavy vectors $Z_{\mu}$ and $V^0_{\mu}$ are more complicated,
but can be approximated for $g^2,g^{\prime \, 2}\ll g_V^2$ as
\beqs
Z_{\mu} &\simeq & \frac{g}{\sqrt{g^2+g^{\prime\,2}}}W^3_{\mu}\,-\,\frac{g^{\prime\,}}{\sqrt{g^2+g^{\prime\,2}}}B_{\mu}-\frac{g^{\prime}}{g_V}
\frac{g^{\prime\,}}{\sqrt{g^2+g^{\prime\,2}}}V^3_{\mu}\,,\\
V^0_{\mu}&\simeq&V^3_{\mu}\,-\,\frac{g^{\prime}}{g_V} B_{\mu}\,,
\eeqs
where we have neglected terms of ${\cal O}(g^{2}/g_V^2)$ and of ${\cal O}(g^{\prime \, 2}/g_V^2)$.

The electric charge is given by
\beq
e\equiv\frac{gg^{\prime}g_V}{\sqrt{g^2g^{\prime\,2}+g^{\prime\,2}g_V^2 +g^{2}g_V^{2}}}\,.
\eeq
 Then, with the electroweak $U(1)$ gauge coupling defined to be
\beqs
g^{\prime}_{\mathrm{SM}}&=&\frac{g^{\prime}g_V}{\sqrt{g_V^2+g^{\prime\,2}}}\,,
\eeqs
we have the conventional relation $e = g^{\prime}_{\mathrm{SM}}g_{\mathrm{SM}}/ \sqrt{g^{\prime\,2}_{\mathrm{SM}} + g^{2}_{\mathrm{SM}}}$.
From here on, we take $g_V^2 \gg g^2,g^{\prime \, 2}$ in all expressions. Then, $g^{\prime}\simeq g^{\prime}_{\mathrm{SM}} \simeq 0.36$.

The masses of the two physical scalars are obtained by writing $\Phi=(f+h) \mathbb{I}_2$ and $\Phi_V=(f_V+H) \mathbb{I}_2$, and are given by
\beqs
m_h^2&=&2\lambda f^2\,,\,\,\,\,\,
m_H^2\,=\,2\lambda_V f_V^2\,.
\eeqs
For $m_h=125$ GeV we have $\lambda\simeq 0.13$, while  $m_H=750$ GeV and $f_V \lsim f$ require
 $\lambda_V  \gsim  9/2$. We make this estimate more precise when considering the properties of the $750$ GeV scalar.

%%%%%%%%%%%%%%%%%%%%%%%%
\section{Precision Electroweak and Higgs physics}
\label{Sec:precision}
%%%%%%%%%%%%%%%%%%%%%%%%

The new sector of our model communicates with the SM fields only through the
$B_{\mu}$ gauge field with strength $g^{\prime}$. The new particles are heavy relative to the SM
particles by virtue of the fact that $g_V^2 \gg g^2, g^{\prime \, 2}$ and $\lambda_V \gg \lambda$.
On the other hand, with masses below $1$ TeV, the new particles are light
enough to raise concerns about their impact on electroweak precision studies.  Deviations from the SM
 are captured in a set of electroweak precision  parameters~\cite{PT,Barbieri}, related to a set of local operators, which
 arise from integrating out the heavy physics, at tree-level and loop-level. The local operators are to be used
at tree level along with loop effects arising from SM particles. The leading parameters and associated
operators through dimension-six are tabulated in~\cite{Barbieri}.

At low energies the new physics in our model leads only to local, gauge invariant operators constructed from
the $B_{\mu}$ field.  Here we focus on operators contributing to the electroweak precision observables.
The dimension-four operator is Tr $B_{\mu\nu} B^{\mu\nu}$, re-scaling a term already in
the Lagrangian. At tree level the single dimension-six operator, arising  from mixing with the heavy sector  is
$\Tr (\partial_{\sigma} B_{\mu\nu} ) (\partial^{\sigma} B^{\mu\nu})$. It affects only the neutral-vector-boson
two-point function and  corresponds directly to the $Y$ parameter of Ref.~\cite{Barbieri}. A  simple tree-level calculation in the
limit $g_V^2 \gg g^2, g^{\prime \, 2}$ yields
\beqs
Y \,\simeq\,\frac{g^{\prime\,2}g^2f^2}{g_V^{4}f_V^2} \,=\,  \frac{g^{\prime\,2}M_W^2}{g_V^{2}M_{V^{\pm}}^2}
\,,
\eeqs
exhibiting both mass and mixing suppression.  According to~\cite{Barbieri}, the bound on $Y$ is
$|Y|<0.0006$ (at $1\sigma$ c.~l.). This translates into the mild bound $g_V M_{V^{\pm}}\gsim 1$ TeV, which
is well respected by our model. In the limit $g_V \rightarrow {\cal O}(4 \pi)$, this estimate becomes very
small, but should be regarded as order-of-magnitude since higher order corrections can become important.

The other precision parameters of Ref.~\cite{Barbieri}, in particular the $\hat{S}$ parameter, are not generated in our model until we
enlarge it to include new Lagrangian terms suppressed by  a very high-scale (see Section~\ref{Sec:symmetry}).
We will take these terms to violate the $U(1)_V$ symmetry in order to allow for decay of the $V^{\pm}_\mu$ bosons.
They will make very small contributions to the precision electroweak parameters.\footnote{With the other precision
parameters vanishing, the $Y$ parameter is related simply
 to the $\rho$ parameter via $\Delta \rho \equiv
\rho - 1 = Y g_{}^{\prime\,2}/ g_{}^2$~\cite{Barbieri}. Thus, in the limit $g_V^2 \gg g^2, g^{\prime \, 2}$, at the tree level we would have
\beqs
\Delta \rho \equiv \frac{M_W^2}{M_Z^2} \frac{g^{\prime\,2}_{\mathrm{}} + g^{2}_{\mathrm{}}}{g^{2}_{\mathrm{}}} - 1
\approx \frac{g^{\prime\,4}f^2}{g_V^{4}f_V^2} + \cdots\,.
\eeqs
}

Some of  the couplings of the Higgs particle $h$ have been measured
precisely enough that they can also be a source of concern for models of new physics ~\cite{precisionh,ABIP}.
In our model,  because of the $U(1)_V$ symmetry, the couplings of $h$ to the fermions and to charged $W$ bosons are identical to those of the SM.
Hence there are no tree-level corrections to the process $h\rightarrow WW$ and no sizeable corrections to
the loop-induced $h\rightarrow \gamma\gamma$.
However, the tree-level coupling of $h$ to $Z_\mu$ bosons is modified by the effect of mixing in the neutral sector between the photon, $Z_\mu$ and $V^0_\mu$.
The experimental bounds coming from the $h\rightarrow ZZ^*$ process~\cite{precisionh},
 can be written in terms of the signal significance as $\mu_{ZZ}=1.31^{+0.27}_{-0.24}$ ($1\sigma$ c.l.). 
We exhibit the coupling by writing $\Phi=(f+h) \mathbb{I}_2$, giving an interaction of the form
\beqs
{\cal L}_{h}&=&\frac{h}{f}\sum_{i,j}V_i^\mu (M_i a_{hij}M_j )V_{j \mu}\,,
\eeqs
where $V_i=(Z,V^0)$ and $M_i=(M_Z,M_{V^0})$. In the large-$g_V$ limit, the $2 \times 2$ interaction matrix reads:
\renewcommand{\arraystretch}{1.5}
\beqs
a_{hij}&=&
\left(
\begin{array}{cc}
 1\,-\, \frac{g_{\mathrm{}}^{\prime\,4} f^2 }{g_V^4 f_{V}^2 }
 &\frac{g_{\mathrm{}}^{\prime\,2} f}{g_V^2 f_{V}}
   \\
 \frac{g_{\mathrm{}}^{\prime\,2} f }{g_V^2 f_{V}}
 & \frac{g_{\mathrm{}}^{\prime\,4} f^2 }{g_V^4 f_{V}^2}
\end{array}
\right)\,,
\eeqs
where again we have kept only the leading terms in the $1/g_V^2$ expansion.
As the $h \rightarrow ZZ^*$ branching ratio is small, and the  $h$ production mechanism is unaffected,
we can approximate $\mu_{ZZ}\simeq a_{hZZ}^2\mu_{ZZ}^{\mathrm{SM}}$.
Together with the experimental bound quoted earlier, this allows us to conclude that at the $3\sigma$ level
\beqs
a_{hZZ}^2 \simeq\,1-2\frac{g^{\prime\,4}f^2}{g_V^{4} f_V^2}\,\gsim\, 0.6\,.
\eeqs
Because of the small  deviation from the SM, suppressed by factors ${\cal O}(g^{\prime\,4}/g_V^4)$
(the same order as the precision $Y$ parameter), this constraint is easily satisfied.

%%%%%%%%%%%%%%%%%%%%%%%%%%
\section{Heavy Scalar Resonance}
\label{Sec:H}
%%%%%%%%%%%%%%%%%%%%%%%%%%

We identify $H$ with the heavy particle seen in diphoton searches at $750$ GeV~\cite{diphoton}.
We borrow the conventions of~\cite{cern}, and write the production cross-section as
\beqs
\sigma(pp\rightarrow H \rightarrow \gamma\gamma)&=&\frac{1}{M_H\Gamma_H s}\left[\frac{}{}\sum_i C_{ij} \Gamma(H\rightarrow ij)\right]\Gamma(H\rightarrow \gamma\gamma)\,,\label{Eq:sigma}
\eeqs
where $\sqrt{s}=13\,{\rm TeV}$, and where the sum is over all the partons in the initial state, including the photon.
The partonic luminosities $C_{ij}$ 
are discussed in the Appendix. In order to reproduce the experimental excesses, it may be necessary 
for cross sections to be  as large as 
$\sigma\simeq (4.8\pm 2.1)\, \mathrm{fb}$ (CMS)
and $\sigma\simeq (5.5\pm 1.5)\, \mathrm{fb}$  (ATLAS)~\cite{Strumia}.
We estimate the parameters of our model required to provide cross sections in this range, 
or somewhat smaller given that the signals are preliminary. 

We assume that the masses of the $V^i$ bosons are large enough to forbid decays of $H$ into
final states involving these bosons.
 As noted earlier, this will be achieved by taking $f_V \lsim f$ and
$ g_V^2 \gg g^2, g^{\prime \, 2}$.
The quantum number assignments are such that $H$ cannot decay directly to two fermions, as
only $\Phi$ enters the Yukawa couplings and there is no mixing between $H$ and $h$.
The absence of mixing in the charged vector-boson sector, a consequence of the $U(1)_V$ symmetry of the
dimension-four Lagrangian,
 ensures that $H$ cannot decay to $W$-boson pairs.\footnote{Both of these conclusions are altered by small mixing effects between $h$ and $H$ and by even smaller $U(1)_V$ symmetry breaking couplings, discussed later in the paper. Both of these give highly suppressed contributions to the width of the $H$.}
  The requirement $M_{V^0}=g_V f_V/2\gsim 660$ GeV
 forbids the decay $H \rightarrow V^{0}Z^{0}$. Thus, the only allowed decays are $H\rightarrow \gamma\gamma,\gamma Z, ZZ$.  The latter can proceed at the tree level while all three can proceed at the quantum level
via loops of charged $V^\pm_{\mu}$ vector bosons. The production process is
dominated by photon fusion via the same loops.

%%%%%%%%%%%%%%%%%%%%%%%%%%
\subsection{Tree-level ZZ Decay}

As with the Higgs scalar $h$, we write the couplings of  $H$ to the neutral vector-boson
mass eigenstates
by writing $\Phi_V= (f_V+H ) \mathbb{I}_2$, so that the Lagrangian contains a coupling
\beqs
{\cal L}_{H}&=&\frac{H}{f_V}\sum_{i,j}V_i^\mu (M_i a_{Hij}M_j) V_{j \mu}\,,
\eeqs
where $V_i=(Z,V^0)$ and $M_i=(M_Z,M_{V^0})$. For large $g_V$, the mixing matrix reads:
\beqs
a_{Hij}&=&
\left(
\begin{array}{cc}
 \, \frac{g_{\mathrm{}}^{\prime\,4} f^2 }{g_V^4 f_V^2 } &
- \frac{g_{\mathrm{}}^{\prime\,2} f }{g_V^2 f_V}
   \\
- \frac{g_{\mathrm{}}^{\prime\,2} f }{g_V^2 f_V}
 & 1-\frac{g_{\mathrm{}}^{\prime\,4} f^2 }{g_V^4 f_V^2}
\end{array}
\right)\,.
\eeqs
By unitarity,  $a_{h}+a_{H}=\mathbb{I}_2$.

The decay rate, adapted from the analogous SM decay of a heavy Higgs particle~\cite{Djouadi:2005gi},  reads
\beqs
\Gamma(H\rightarrow ZZ)&=&a_{HZZ}^2\frac{m_H^3}{32\pi f_V^2}\sqrt{1-4x}\left(\frac{}{}1-4x+12x^2\right)\,,
\eeqs
where $x=M_Z^2/m_H^2$. Here, with $x\ll 1$ and with
\beqs
a_{HZZ} \simeq\,\frac{g_{\mathrm{}}^{\prime\,4}M_W^2 }{g^2g_V^2M_{V^{\pm}}^2}\,,
\eeqs
 we have
\beqs
\Gamma^{\mathrm{tree}}&\simeq &a_{HZZ}^2\frac{m_H^3}{32\pi f_V^2}\,=\,
\frac{g_{\mathrm{}}^{\prime\,8}M_W^4 m_H^3}{32 \pi\,g^4g_V^4 M_{V^{\pm}}^4f_V^2}\,.
\eeqs
For $g_V^2 \gg g^2, g^{\prime \, 2}$, the requisite mixing renders the width very small.
As we discuss in Section~\ref{Sec:symmetry}, a comparable contribution to the amplitude will come from the small mixing between $H$ and $h$ arising at the loop level.  Both are dominated by loop contributions that do not suffer from such strong mixing suppression.

%%%%%%%%%%%%%%%%%%%%%%%%%%
\subsection{Loop-induced Decays}

The coupling of $H$ to two photons is controlled by a dimension-five operator generated by loops of charged particles.
Because of the simple structure of the model, the only one-loop contribution comes from the
 $V^{\pm}_\mu$, as no fermions couple to $H$. Adapting the decay rate of the Higgs
boson to photons in the SM to the present case~\cite{Djouadi:2005gi}, we have
\beqs
\Gamma(H\rightarrow \gamma\gamma)&=&\left(\frac{v_W}{f_V}\right)^2\frac{G_F\alpha^2m_H^3}{128\sqrt{2}\pi^3}\left|
A_W\left(\frac{m_H^2}{4M_{V^{\pm}}^2}\right)\right|^2\,,
\eeqs
where
\beqs
A_W(x)&=&-\frac{1}{x^2}\left(\frac{}{}2x^2+3x+3(2x-1){\rm arcsin}^2\sqrt{x}\right)\,.
\eeqs
The appearance of $v_W$  is for convenience. It is cancelled by the factor $G_F$, leaving only scales associated with the new sector.
In the limit $x \ll 1$, relevant to the present case, one finds $A_W\left(\frac{m_H^2}{4M_{V^{\pm}}^2}\right)\rightarrow -7$, and we have
\beqs
\Gamma(H\rightarrow \gamma\gamma)&\simeq& 3\,\,{\rm MeV}\,\, \left(\frac{v_W}{f_V}\right)^2\,.
\eeqs

The loop-induced decays to $ZZ$ and $Z\gamma$, in the limit $M_Z\ll m_H$,
are obtained in the same manner. Neglecting the tree-level contribution to the decay to $ZZ$, the one-loop expressions (for $g_V^2 \gg g^2,g^{\prime \, 2}$) are
\beqs
\Gamma(H\rightarrow Z\gamma)&=&2\left(\frac{g^{\prime}}{g}\right)^2\Gamma(H\rightarrow \gamma\gamma)\,\simeq\,0.61\,\,\Gamma(H\rightarrow \gamma\gamma)\,,\\
\Gamma(H\rightarrow ZZ)&=&\left(\frac{g^{\prime}}{g}\right)^4\Gamma(H\rightarrow \gamma\gamma)\,\simeq\,0.09\,\Gamma(H\rightarrow \gamma\gamma)\,.
\eeqs
The one-loop two-body decay widths together yield
\beqs
\Gamma^{\mathrm{tot}}&\simeq&\frac{(g^2+g^{\prime\,2})^2}{g^4}\,\Gamma(H\rightarrow \gamma\gamma)\,\simeq\,1.71\,\,\Gamma(H\rightarrow \gamma\gamma)\, \simeq\, 5\,\,{\rm MeV}\,\, \left(\frac{v_W}{f_V}\right)^2\,,
\label{Eq:totrate}
\eeqs
giving a relatively narrow width and a large diphoton branching ratio  $\mathrm{BR}_{\gamma \gamma}$ in the relevant range of parameter space.

%%%%%%%%%%%%%%%%%%%%%%%%%%
\subsection{$H$ Production and Parameter Estimates}
\label{Sec:HP}

Production proceeds via photon fusion.  The expression for the $\gamma \gamma$ production cross section is
\beqs
\sigma(pp\rightarrow H \rightarrow \gamma\gamma)&=&
\frac{\mathrm{BR}_{\gamma\gamma}}{M_H s}C_{\gamma\gamma}\Gamma(H\rightarrow \gamma\gamma)\,
\simeq\,0.5\,\,{\rm fb}\,\,\left(\frac{v_W}{f_V}\right)^2\,\mathrm{BR}_{\gamma\gamma}\,,
\eeqs
where $\sqrt{s}= 13$ TeV.

With the large value of BR$_{\gamma\gamma}$ implied by Eq.~(\ref{Eq:totrate}), this expression can approach the level indicated by the experimental signals in diphoton searches, but only if $f_V$ is no larger than $v_W$. If the estimates following Eq.~(\ref{Eq:sigma}) are accurate, $f_V$ might have to be as low as $ \sim100$ GeV, giving a production cross section times branching ratio of $\sim 2$ fb.  On the other hand, if smaller cross sections are eventually revealed, then we could have $f_V \sim v_W$.  These estimates are affected
by large uncertainties in the partonic luminosity $C_{\gamma\gamma}$ (see the Appendix). Because the total width is small, our model can accommodate the hints of new resonant production only in terms of a single, narrow-width resonance.

If $f_V$ must be as low as $100$ GeV, then $g_V^2 = 4M_V^2 /f_V^2 \rightarrow {\cal O}(16\pi^2)$ and $ \lambda_V = 2M_{H}^2 /f_{V}^2 \rightarrow {\cal O}(16\pi^2)$. 
The new sector in our model becomes strongly coupled, raising the question of whether it truly functions as an EFT, capturing
 all the relevant degrees of freedom and allowing for reliable approximate computations.
If not, our one-loop estimates for the decay widths and branching ratios of the new scalar $H$ should be regarded as order-of-magnitude estimates. The same is true of our estimates of the precision electroweak parameters, but they were comfortably small due to mixing and mass suppression.

If it develops that the production cross section  Eq.~(\ref{Eq:sigma}) is small enough to allow $f_V \sim v_W$,  then $g_V^{2},\lambda_V < (4\pi)^2$, and perturbation theory could
become reliable. The cutoff on the EFT, to be discussed in Section~\ref{Sec:symmetry}, would be somewhat above
the masses of the new scalar and vectors, perhaps in the few-TeV range. Our model would
function as a bonafide EFT over a limited energy range. Either way, the model provides a plausible
explanation for a $750$ GeV resonance with a large branching ratio to photons.

%%%%%%%%%%%%%%%%%%%%%%%%%%
\section{Vector resonances}
\label{Sec:V}
%%%%%%%%%%%%%%%%%%%%%%%%%%

With parameters in the range described above, $V^0_\mu$ and $V^{\pm}_\mu$ are approximately degenerate and have masses in the range of the
diphoton resonance $H$.

%%%%%%%%%%%%%%%%%%%%%%%%%%
\subsection{Neutral Vector}

The neutral vector is produced and decays dominantly via tree-level processes, induced by the mixing
between $B_\mu$, $W^3_\mu$ and $V^0_\mu$.
We write the general LHC production cross-section as in~\cite{cern}:
\beqs
\sigma(pp\rightarrow V^0\rightarrow X)&=&\frac{3}{M_{V^0}\Gamma_{V^{0}} s}\left[\frac{}{}\sum_{ij}C_{ij}\Gamma(V^{0}\rightarrow ij)\right]\,\Gamma(V^{0}\rightarrow X)\,.
\eeqs
The relevant parton luminosities $C_{ij}$  are the quark-antiquark ones (see Appendix A).

The dominant contributions to the decay rates come from two-body final states, which for kinematical reasons involve only SM particles.
Because of the large mass of the $V^0_\mu$, we approximate the rates by ignoring the masses of the final state particles.

We write the effective couplings to the fermions as
\beqs
{\cal L}&=&\frac{g^{\prime}}{g_V}\,g^{\prime} V^{0}_{\mu}J_{Y}^{\mu}\,
\,+\, {\cal O}\left( \frac{1}{g_V^2} \right)
\eeqs
where $J_{Y}^{\mu}$ is the $U(1)_Y$ current made of fermions.
We then obtain:
\beqs
\Gamma(V^0\rightarrow u\bar{u})&=&\frac{17}{288\pi}\,\frac{g^{\prime\,4}}{g_V^2}M_{V^0}\,,\\
\Gamma(V^0\rightarrow d\bar{d})&=&\frac{5}{288\pi}\,\frac{g^{\prime\,4}}{g_V^2}M_{V^0}\,,\\
\Gamma(V^0\rightarrow e^+e^-)&=&\frac{5}{96\pi}\,\frac{g^{\prime\,4}}{g_V^2}M_{V^0}\,,\\
\Gamma(V^0\rightarrow \nu_e\bar{\nu}_e)&=&\frac{1}{96\pi}\,\frac{g^{\prime\,4}}{g_V^2}M_{V^0}\,.
\eeqs
The total width for decay to three families of fermions, treating them all as massless, is
\beqs
\Gamma(V^0\rightarrow \psi\bar{\psi})&=&\frac{5}{12\pi}\frac{g^{\prime\,4}}{g_V^2}M_{V^0}\,.
\eeqs

The decay to SM bosons ($WW$ and $Z h$) yields a smaller contribution to the total width:
\beqs
\Gamma(V^0\rightarrow W^+W^-)\simeq \Gamma(V^0\rightarrow Z h)\simeq \frac{1}{192\pi}\frac{g^{\prime\,4}}{g_V^2}M_{V^0}\,.
\eeqs
We conclude that the branching ratio to electron-positron is
$\mathrm{BR}(V^0\rightarrow e^+e^-)\simeq\frac{1}{8}$. Taking $g^{\prime}\simeq 0.36$, 
$g_V\lsim 4\pi$ and making the representative choice $M_{V^0}\simeq 700$ GeV, we find that the predicted LHC production cross-sections for $e^+ e^-$ and $\mu^+ \mu^-$ (collectively denoted by $\ell^+ \ell^-$) are:
\beqs
\sigma(pp\rightarrow V^0\rightarrow \ell^+ \ell^-)_{8 \mathrm{TeV}}&\simeq&(21000 \,\mathrm{fb})\,\frac{g^{\prime\,4}}{g_V^2}\,\gsim 2\,{\mathrm{fb}}\,,\\
\sigma(pp\rightarrow V^0\rightarrow \ell^+\ell^-)_{13 \mathrm{TeV}}&\simeq&(53000 \,\mathrm{fb})\,\frac{g^{\prime\,4}}{g_V^2}\,\gsim\,6\,{\mathrm{fb}}\,,
\eeqs
at $\sqrt{s}=8$ TeV and $13$ TeV, respectively. 
The explicit dependence on $M_{V^{0}}$ drops from the cross-section, 
but it re-enters through the PDF, evaluated at the scale $M_{V^0}$.
For indicative choices of parameters relevant to our study,
the width is $\Gamma\simeq {\cal O}(10$ MeV).

From the LHC run at $8$ TeV, the upper bound on the above cross section from ATLAS~\cite{dilepton8} for $M_{V^0}\simeq 700$ GeV
is  $\sigma(pp\rightarrow V^0\rightarrow \ell^+\ell^-)\lsim 2$ fb. In  the $13$ TeV case~\cite{dilepton13} it is $\sigma(pp\rightarrow 
V^0\rightarrow \ell^+\ell^-)\lsim 7$ fb \footnote{The bounds quoted here are computed from the experimental data by assuming that the candidate heavy vector has the branching ratios of an SM $Z$ boson but with a mass of 700 GeV.}.
In the limit $g_V\rightarrow 4\pi$, the bounds appear to be satisfied, although the estimates should be viewed as order-of-magnitude
due to strong coupling as discussed in Section~\ref{Sec:HP}.
For the case $g_V^2<(4\pi)^2$, corresponding to $f_V\sim v_W$, the estimates
appear to be at or somewhat above the published experimental bounds.  However, the combination of experimental error and uncertainties in our theoretical estimates associated, for example, with the $M_{V^0}$-dependence of the parton distribution functions, leaves open the possibility that our predictions remain compatible with current upper bounds.

These estimates offer the exciting possibility that with higher integrated luminosity the production
and decay of the $V^0_\mu$ should be observable. It should be possible to test the viability
of the current model by looking for di-lepton excesses with an invariant mass comparable to $m_H$ at the LHC.

%%%%%%%%%%%%%%%%%%%%%%%%%%
\subsection{Charged Vectors}

In the charged sector, the presence of the $g^{\prime}$ coupling does not lead to mixing with the SM gauge bosons.
The charged $V_{\mu}^{\pm}$ is stable due to the $U(1)_V$ symmetry necessarily present in the dimension-four Lagrangian.
When higher-dimension terms are added, however, this symmetry is easily broken. A
simple example is the dimension-six operator
\beq
 \frac{1}{\Lambda_V^2} \Tr \left[\frac{}{}(D_{\mu}\Phi) \Phi_V
 (D^{\mu}\Phi_V)^{\dagger}\Phi^{\dagger}\right]\,,
\label{Eq:higher}
\eeq
which can lead, for example, to the decay $V^{\pm} \rightarrow W^{\pm} h$.
The scale $\Lambda_V$ at which the $U(1)_V$ symmetry is broken
must be large enough so that operators of this type lead to
only small corrections
to the precision quantities discussed in Section~\ref{Sec:precision}. Conversely, the scale must
be small enough to ensure that
the $V^{\pm}$ decays rapidly enough to avoid direct detection.
The broad range $1$ TeV $\ll \Lambda_V \lsim 1000$ TeV will allow both
constraints to be easily satisfied.

The production cross-section for the $V^{\pm}_{\mu}$ is dominated by pair production mediated by a virtual $\gamma_\mu$, $Z_\mu$ or $V^0_\mu$.
The cross section is suppressed with respect to the $V^0_\mu$ production cross-section, 
in part because the parton luminosities are
smaller at the higher energies needed to produce on-shell pairs $V^\pm_\mu$.
Also the final states of the decays are more difficult to reconstruct. We conclude that these processes are less likely to yield important future signals
than processes involving the $V^{0}_\mu$.

%%%%%%%%%%%%%%%%%%%%%%%%%%
\section{Symmetry and fine tuning}
\label{Sec:symmetry}
%%%%%%%%%%%%%%%%%%%%%%%%%%

Our model has the following features.

\begin{itemize}

\item At the dimension-four level, before gauging and neglecting Yukawa couplings, the model possesses a global  $SU(2)^4$ symmetry. (The $SU(2)_R$, symmetry in Table~\ref{Fig:3sites} can act independently on the new sector and the SM sector.)  This symmetry is present even when including all possible terms in the dimension-four potential, including the sector-coupling term $\Tr(\Phi\Phi^{\dagger}) \Tr(\Phi_V\Phi_V^{\dagger}) = 2 \Tr(\Phi\Phi_V\Phi_V^{\dagger}\Phi^{\dagger})$, neglected so far.

 \item    The $SU(2)^4$ symmetry is explicitly broken
by the gauging of an $SU(2)_L\times U(1)_Y\times SU(2)_V$ subgroup, preserving an additional, accidental, global  $U(1)_V$ symmetry. At the dimension-four level, communication between the new sector and the SM fields
is via the gauging of the hypercharge $Y$, controlled by the small coupling $g^{\prime}$.

\item The fermionic field content, charge assignments  and dimension-four
Yukawa interactions are chosen in such a way so as to preserve the same symmetry, and to reproduce the SM at low energies.

\item The $U(1)_V$ accidental symmetry, which stabilizes the $V^{\pm}_{\mu}$, is broken only by higher-dimension operators 
such as the one  in~(\ref{Eq:higher}),
suppressed by a very high scale $\Lambda_V$.

\item The self-coupling of $\Phi_V$ and the gauge coupling of $SU(2)_V$ are much larger than the electroweak couplings, possibly reaching the non-perturbative limit $\lambda_V , g_V^2 \rightarrow (4 \pi)^2 $.

\end{itemize}

We next discuss corrections to the scalar potential, which has not included the mixing term
between $\Phi$ and $\Phi_V$, and the issue of fine tuning.
 The one-loop effective potential, arising from the gauge bosons only, can be
 written in terms of the cutoff scale $\Lambda$ 
 and renormalization scale $\mu$ as
\beqs
V_1&=&
\frac{3 \Lambda ^2}{256 \pi ^2} \left[\frac{}{}\left(3 g^2+g^{\prime\,2}\right)\Tr\Phi^{\dagger}\Phi
+\left(g^{\prime\,2}+3 g_V^2\right)\Tr\Phi^{\dagger}_V\Phi_V
   \right]
  +\\
&&\nonumber
+\frac{3}{4096 \pi ^2}\left[  \left(3 g^4+2 g^2 g^{\prime\,2}+g^{\prime\,4}\right)(\Tr\Phi^{\dagger}\Phi)^2
+2 g^{\prime\,4}(\Tr\Phi^{\dagger}\Phi)(\Tr \Phi_V\Phi_V^{\dagger} )
  \right.\nonumber\\
  &&\left.\frac{}{}
   + \left(g^{\prime\,4}+2 g^{\prime\,2} g_V^2+3
   g_V^4\right)(\Tr\Phi_V\Phi^{\dagger}_V)^2\right]\ln\frac{\mu^2}{\Lambda^2}\,+\cdots\,,\nonumber
\eeqs
In addition, there are $\Lambda^2$  and $\ln \Lambda^2$ terms arising from the scalar self-couplings and from
the fermion Yukawa couplings. 
We omit them for simplicity.

Consider first the terms quadratic in the cutoff, which contribute to the masses of $h$ and $H$. The $\Tr(\Phi\Phi^{\dagger})$
term (including additional contributions from the Higgs self coupling and top-quark Yukawa coupling)
will not require substantial fine tuning as long as $\Lambda$ is less than a few TeV. The $\Tr(\Phi_V\Phi_V^{\dagger})$
term has a contribution proportional to $g_V^2$,  as well as to $\lambda_V$ (not shown).
Again, the cutoff can be no larger than a few TeV to avoid substantial fine tuning of the mass $m_H$. With $g_V$ and $\lambda_V$ approaching the strong coupling limit, this estimate is only order-of-magnitude. Interestingly, with the top loop dominating the $\Tr(\Phi\Phi^{\dagger})$ term, it has the opposite sign of the one-loop $\Tr(\Phi_V \Phi_V^{\dagger})$ term.

At the $\ln {\Lambda^2}$ level, the $(\Tr\Phi_V\Phi^{\dagger}_V)^2$  term can be of the same order of magnitude
as the corresponding tree-level term if $g_V^2$ and $\lambda_V$ approach the strong-coupling level.
There is no fine-tuning issue here, although once again the estimates are order-of-magnitude in the strong coupling limit. The $(\Tr \Phi\Phi^{\dagger})^2$ term is smaller.

The mixing term, which is proportional to only $g^{\prime\,4}$, is the smallest of the ln$\Lambda^2$ terms since $g^{\prime\,4} \ll g^4 \ll g_V^4$. Although it breaks no symmetries and is dimension-four, it leads to only small mixing effects between the Higgs field $h$ and the heavy scalar $H$, and can consistently be neglected to leading approximation.

 Finally, we note again that each of the allowed dimension-four terms in our Lagrangian, (including the mixing term) necessarily respects the global $U(1)_V$ symmetry.
 Since this accidental symmetry stabilizes the $V^{\pm}_\mu$ bosons, we have invoked new $U(1)_V$-breaking interactions
 at a very high scale $\Lambda_V\gg \Lambda$, described by higher-dimension operators,
 to allow them to decay. An example is given in Eq.~(\ref{Eq:higher}).
 These higher-dimension operators can contribute to each of the electroweak parameters,
 but since the associated scale is very high, their contribution will be very small.

%%%%%%%%%%%%%%%%%%%%%%%%%%
\section{Summary and Outlook}
\label{Sec:discussion}
%%%%%%%%%%%%%%%%%%%%%%%%%%

We have proposed a simple model describing   the CMS and ATLAS diphoton excess at $750$ GeV. 
The new resonance is interpreted as a narrow scalar particle $H$
associated with the mechanism for the spontaneous breaking of a new $SU(2)_V$ gauge symmetry.
Thus, the model predicts the existence of three heavy vectors $V^i_{\mu}$. Their masses are required to be large enough to avoid direct decay of $H$ into final states including the new vectors.

The model  yields a ${\cal O}(1 \,\mathrm{fb})$ cross-section at
the LHC for the resonant production of $H$ via photon fusion with the coupling of $H$ to photons induced by loops of charged vector bosons. To make the coupling to photons strong enough to provide a sufficiently large production cross section, $f_V$ can be taken no larger than the electroweak scale. It is therefore necessary to have a large gauge coupling $g_V$, so as to make the new vectors sufficiently massive.

All the indirect bounds from precision electroweak and Higgs physics are satisfied despite the comparatively low masses of the new particles and their large self-interactions. 
We have highlighted  the special features of the model leading to this property, as well as the 
intrinsic uncertainties in our estimates.
The model is directly testable, as at least the neutral $V^0_{\mu}$ has a production cross-section and branching ratios large enough to
be detectable via its leptonic decays in higher-luminosity runs of the LHC.

In the event that the signals of $H$ are confirmed, and that hints of a new $V^0_\mu$ resonance emerge, 
two related lines of enquiry are suggested by this study.
First, the rich and model-dependent phenomenology of the charged $V^{\pm}_{\mu}$ should be 
explored further, as it depends sensitively on the $U(1)_V$-violating operators 
responsible for its finite lifetime.
Second, the simplicity of our EFT leaves open many possibilities for its UV completion,
but the relatively strong coupling of its new sector indicates that the completion lies
nearby in energy. Since our new sector is not responsible for electroweak symmetry breaking, we envisage the UV completion to have properties very different from technicolor-like models.
Whatever its nature, we anticipate new states appearing just above the
TeV scale, where they should be experimentally accessible.

\vspace{0.5cm}

\noindent\large{\textbf{Authors' Note (September, 2016)}}

\vspace{0.5cm}

\noindent The initial, 2015 evidence for an excess in the diphoton mass distribution at 750 GeV, by both the ATLAS and CMS collaborations \cite{diphoton}, was not confirmed in the first 2016 LHC run with $(12.2+12.9)$ fb$^{-1}$ \cite{ichep}. If the absence of the resonance persists, the simple effective field theory developed in this paper must be modified if it is to be viable. It can then be developed further, for example by exploring UV completion. Minimally, a simple parameter adjustment can raise the scalar and vector mass scales associated with the new sector to the few-TeV level, placing them beyond current lower bounds, yet not leading to unacceptable fine tuning. In the absence of experimental evidence even in the few-TeV range, the model would have to modified more substantially if fine tuning is to be avoided. An intriguing feature of the model as it stands is a new, automatic, "accidental" U(1) symmetry, which stabilizes the new charged vector bosons. This symmetry can be broken only by higher-dimension operators, leading naturally to a relatively long  lifetime for these particles. 

%%%%%%%%%%%%%%%%%%%%%%%%%%%%%%%%%%%%%%%%
\begin{acknowledgments}

We thank Yang Bai, Luigi Del Debbio and Sarah Demers for helpful discussions.
The work of TA and JI is supported by the U.S. Department of Energy under the contract DE-FG02-92ER-40704.
The work of MP is supported in part by the STFC Consolidated Grant ST/L000369/1.

\end{acknowledgments}
%%%%%%%%%%%%%%%%%%%%%%%%%%%%%%%%%%%%%%%%
\appendix
%%%%%%%%%%%%%%%%%%%%%%%%%%%%
%%%%%%%%%%%%%%%%%%%%%%%%%%
\section{Parton Distribution Functions}
%%%

\begin{figure}[t]
\begin{center}
\begin{picture}(480,130)
\put(10,10){\includegraphics[height=4.2cm]{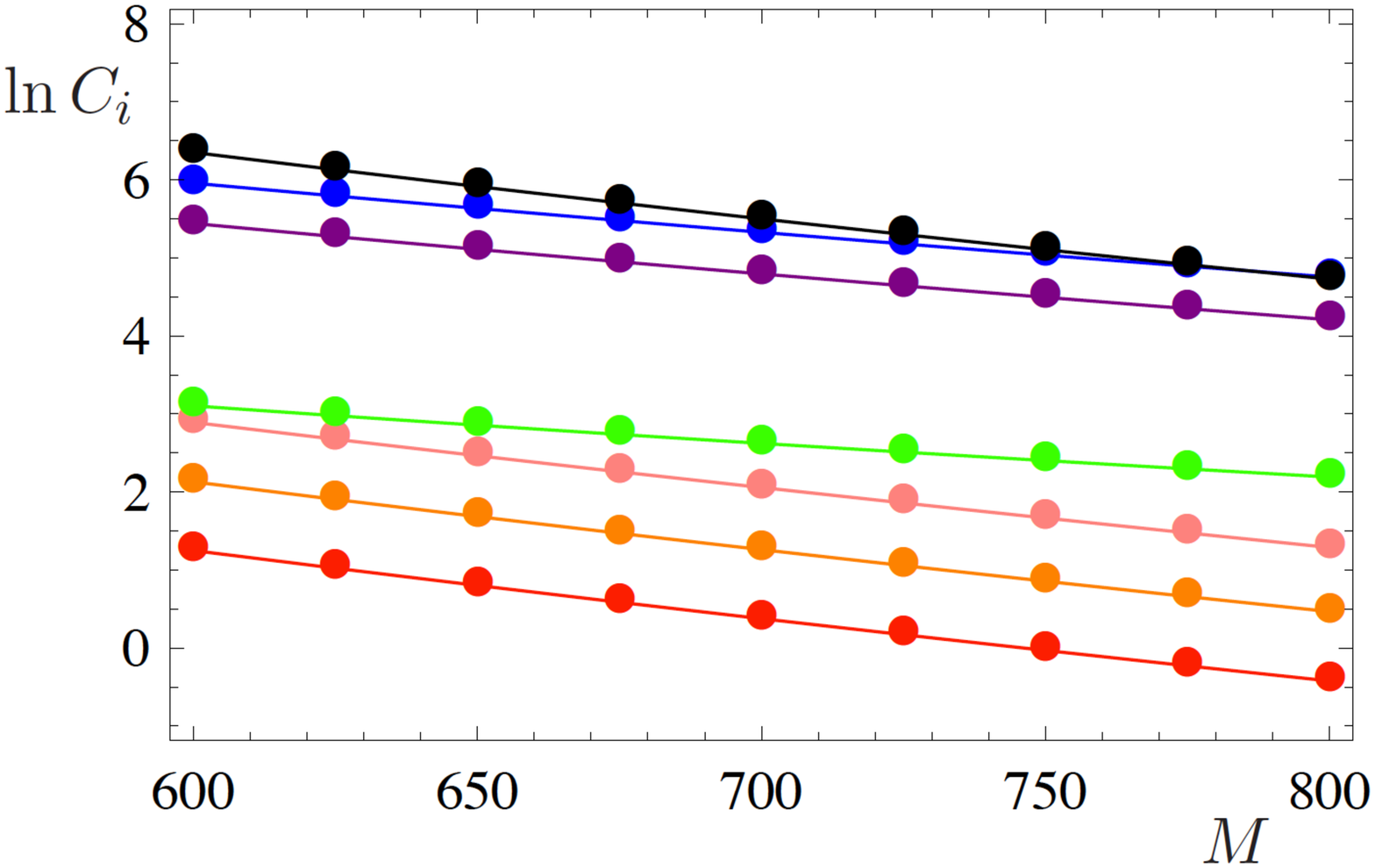}}
\put(225,10){\includegraphics[height=4.2cm]{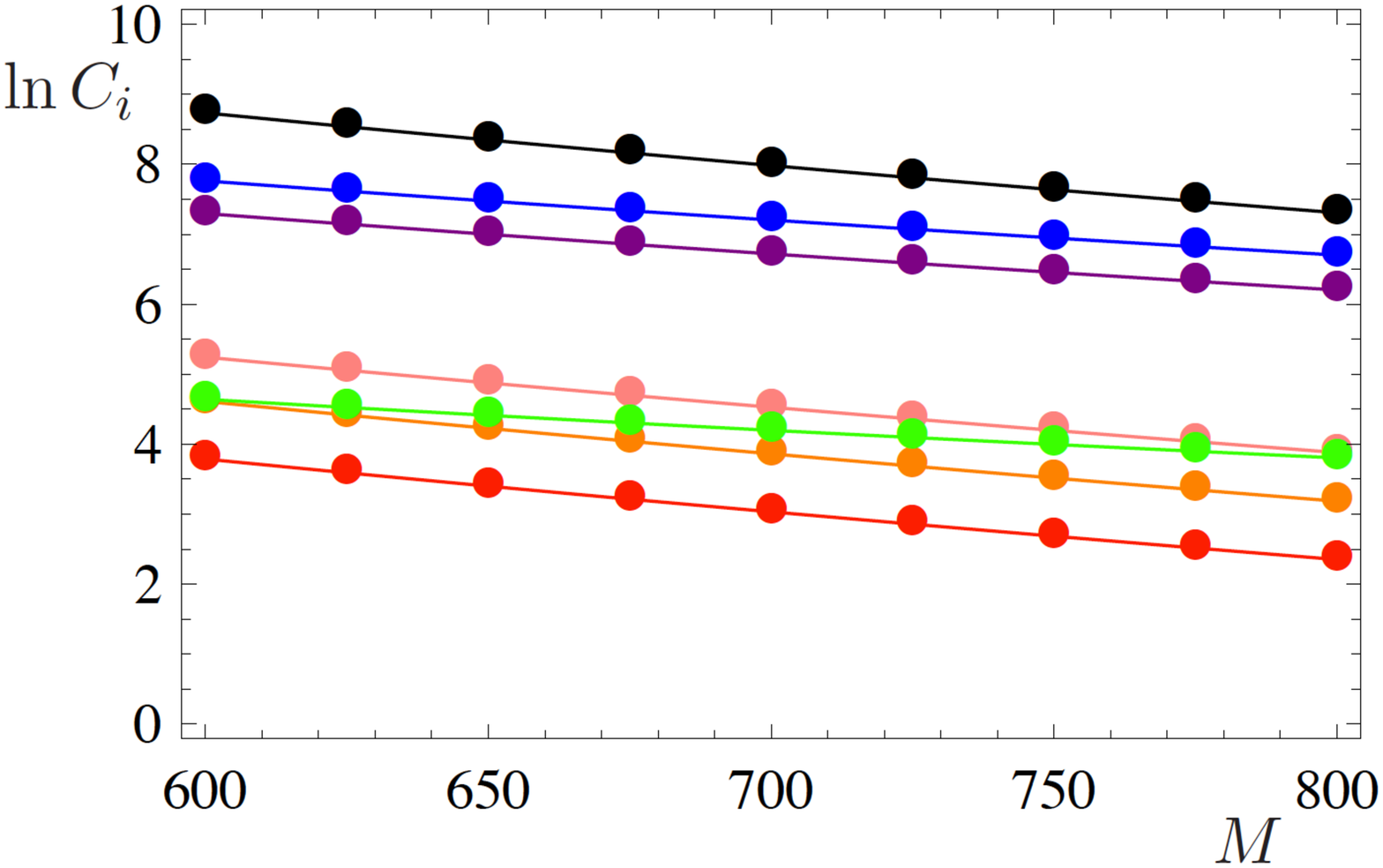}}
\end{picture}
\caption{Central values for the parton luminosities from  NNPDF, as a function of $M$ (in GeV). The left panel has $\sqrt{s}=8$ TeV, the right  $\sqrt{s}=13$ TeV.
We show $C_{b\bar{b}}$ (red),  $C_{c\bar{c}}$ (orange),  $C_{s\bar{s}}$ (pink),  $C_{d\bar{d}}$ (purple),
 $C_{u\bar{u}}$ (blue), $C_{gg}$ (black) and $C_{\gamma\gamma}$ (green). }
\label{Fig:pdf}
\end{center}
\end{figure}

To make predictions for the LHC,
we use the PDFs from NNPDF~\cite{NNPDF,NNPDFMathematica}.
We define the dimensionless parton luminosities with the same conventions as in~\cite{cern}:
\beqs
C_{q\bar{q}}&\equiv&\frac{4\pi^2}{9}\int_{M^2/s}^1\frac{\di x}{x}\left[q(x)\bar{q}\left(\frac{M^2}{x\,s}\right)+\bar{q}(x){q}\left(\frac{M^2}{x\,s}\right)\right]\,,\\
C_{gg}&\equiv&\frac{\pi^2}{8}\int_{M^2/s}^1\frac{\di x}{x} \,g(x) {g}\left(\frac{M^2}{x\,s}\right)\,,\\
C_{\gamma\gamma}&\equiv& {8\pi^2}{}\int_{M^2/s}^1\frac{\di x}{x} \,\gamma(x) {\gamma}\left(\frac{M^2}{x\,s}\right)\,,
\label{photonlum}
\eeqs
where $s$ is the Mandelstam variable and $M$ the relevant scale of the process.
We show in Table~\ref{Fig:PDF} and Fig.~\ref{Fig:pdf} the results for various choices of $s$ and $M$.

We compared these to the parton luminosities  from the MSTW~\cite{MSTW} NLO package,
and found excellent agreement for the quark and gluon luminosities.
The agreement  is at the few percent level, with the exception of the subdominant $b\bar{b}$, where it is at the $10\%$  level.
We also checked that the results we quote agree with~\cite{cern}.

The photon luminosity $C_{\gamma\gamma}$ determines the production cross section for the scalar $H$. It is given in Eq.~(\ref{photonlum}) in terms of the photon PDFs, which have uncertainties ranging from $50 \%$ to $200 \%$ across the domain of the integral. There are also systematic uncertainties in
the evolution of the photon PDFs by NNPDF. Alternative methods exist
for the determination of $C_{\gamma\gamma}$, but a growing body of literature
\cite{MR,CHT,FGR}  indicates that the various methods agree on its
order of magnitude. The combination of statistical and systematic uncertainties imply that the values we used for $C_{\gamma\gamma}$
(obtained from NNPDF) are affected by uncertainties as large as
$C_{\gamma\gamma}$ itself.

\renewcommand{\arraystretch}{1.0}
\begin{table}
\begin{center}
\begin{tabular}{|c|c|ccccccc|
}
\hline\hline
$\sqrt{s}$& $M$& \multicolumn{7}{c|}{{NNPDF}}   \\
(TeV) & (GeV) & $C_{b\bar{b}}$& $C_{c\bar{c}}$& $C_{s\bar{s}}$& $C_{d\bar{d}}$& $C_{u\bar{u}}$& $C_{gg}$
& $C_{\gamma\gamma}$ \\ 
\hline
$8$  & 600. & 3.5 & 8.4 & 18. & 230.6 & 386.4 & 573.1 & 22.3 \\
$8$  & 625. & 2.8 & 6.7 & 14.5 & 194.9 & 328.0 & 459.1 & 19.6 \\
$8$  &  650. & 2.2 & 5.4 & 11.8 & 165.5 & 279.7 & 370.2 & 17.4  \\
$8$  &  675. & 1.8 & 4.3 & 9.6 & 141.1 & 239.5 & 300.2 & 15.4  \\
$8$  &  700. & 1.5 & 3.5 & 7.8 & 120.8 & 205.8 & 244.6 & 13.8 \\
$8$  &  725. & 1.2 & 2.9 & 6.4 & 103.8 & 177.5 & 200.3 & 12.3 \\
$8$  &  750. & 1.0 & 2.4 & 5.3 & 89.5 & 153.6 & 164.7 & 11.0 \\
$8$  &  775. & 0.8 & 1.9 & 4.4 & 77.4 & 133.3 & 136.2 & 9.9 \\
$8$  &  800. & 0.7 & 1.6 & 3.6 & 67.1 & 116.1 & 113.0 & 8.9   \\
\hline
$13$  &   600. & 44.1 & 100.5 & 189.2 & 1475.1 & 2351.8 & 6214.3 & 103.6  \\
$13$  & 625. & 36.2 & 82.7 & 157.0 & 1268.2 & 2031.4 & 5105.5 & 92.2 \\
$13$  &   650. & 29.9 & 68.4 & 131.0 & 1095.6 & 1762.9 & 4220.5 & 82.4 \\
$13$  &  675. & 24.9 & 56.9 & 109.8 & 950.7 & 1536.2 & 3507.8 & 73.9  \\
$13$  &   700. & 20.7 & 47.5 & 92.5 & 828.2 & 1343.9 & 2928.9 & 66.5 \\
$13$  &   725. & 17.4 & 39.9 & 78.2 & 724.2 & 1179.9 & 2457.5 & 60.0 \\
$13$  &   750. & 14.6 & 33.6 & 66.4 & 635.6 & 1039.4 & 2071.6 & 54.3  \\
$13$  &  775. & 12.4 & 28.4 & 56.6 & 559.7 & 918.6 & 1754.0 & 49.3\\
$13$  &  800. & 10.5 & 24.1 & 48.4 & 494.5 & 814.3 & 1491.1 & 44.9  \\
\hline\hline
\end{tabular}
\end{center}
\caption{Parton luminosities obtained  from  the NNPDF package~\cite{NNPDFMathematica}
 and the NNPDF23\_nlo\_as\_0119\_qed.LH grid,
implementing QED and  NLO QCD evolution with $\alpha_s(M_Z)=0.119$ and $\alpha(M_Z)=1/128$.}
\label{Fig:PDF}
\end{table}

%%%%%%%%%%%%%%%%%%%%%%%%%%%%%%%%%%%%%%%%

\end{document}